\documentclass[apjl,a4paper,12pt,useAMS]{emulateapj}
\usepackage{amsmath}
\usepackage{cases}
\usepackage{amssymb}
\usepackage{color}
\usepackage{graphicx}

\begin{document}
\title{Delayed spin-up and persistent shift phenomena of Crab pulsar glitches: two sides of the same coin?}
\author{Wei-Hua~Wang$^{1}$ and Xiao-Ping~Zheng$^{1,2}$}

\altaffiltext{1}{Institute of Astrophysics, Central China Normal
University, Wuhan 430079, China, {wangweihua@mails.ccnu.edu.cn, zhxp@phy.ccnu.edu.cn}}
\altaffiltext{2}{Key Laboratory of Quark and Lepton Physics (Central
China Normal University), Ministry of Education, Wuhan 430079,
China}

\begin{abstract}
Pulsar glitches are sudden increase in their spin frequency, in most cases followed by the
long timescale recovery process. As of this writing, about 546 glitches have been reported
in 188 pulsars, the Crab pulsar is a special one with unique manifestations. This writing
presents a statistic study on post-glitch observables of the Crab pulsar, especially the
delayed spin-up in post-glitch phase and persistent shift in the slow-down rate of the star.
By analyzing the radio data over 45 years, we find that two power law functions respectively
fit the persistent shift and delayed spin-up timescales versus glitch size well, and we find
a linear correlation between the persistent shift and delayed spin-up timescale from
the consistency of the two fitting functions, probably indicating their same physical origin
and may provide a new probe of interior physics of neutron stars.

\end{abstract}

\section{introduction}
Glitch is a phenomenon that interrupts the monotonous spin down of pulsars due to electromagnetic
braking~(Taylor et al. 1993), it is characterized by a sudden increase in spin frequency, generally
accompanied by a increase in spin down rate. The first glitch was discovered in the Vela pulsar in
1969~(Radhakrishnan \& Manchester 1969; Reichely \& Downs 1969), at present, about 546 glitches have
been reported in 188 pulsars~\footnote{http://www.jb.man.ac.uk/pulsar/glitches/gTable.html,
and http://www.atnf.csiro.au/people/pulsar/psrcat/glitchTbl.html.}, the famous Crab and Vela pulsars
are both frequently glitch sources and are daily monitored. A series of models have been proposed
ever since its first discovery, such as crustquake~(Ruderman 1969), corequake~(Pines et al. 1972),
planetary perturbation~(Michel 1970) and magnetospheric instabilities~(Scargle \& Pacini 1971), but
none of these were convincing enough~(Pines et al. 1974). In 1969, Baym et al. proposed the long
timescale in the post-glitch recovery process of the Vela pulsar as a signature of neutron superfluid
in inner neutron star~(Baym et al. 1969). It should be noted that the absence of radiative and pulse
profile changes in Vela glitches seems to support its internal origin. In 1975, Anderson and Itoh
advanced the semina idea that glitches are triggered by sudden unpinning of superfluid vortices from
neutron star crust~(Anderson \& Itoh 1975), resulting in a rapid transfer of angular momentum from
the faster rotating superfluid component to the normal component, besides, as a small portion of
moment of inertia decouples from the normal component while the external torque acting on the pulsar
remains constant in short timescale, the observed spin down rate will thus increase temporarily. Alpar et al.
further developed this into the vortex creep theory~(Alpar et al. 1984), which is now widely
accepted as the standard scenario due to its success in explaining the post-glitch
recovery process.

Within the framework of vortex creep theory, the pinning force and the friction between the crust
and the superfluid component dominate the post-glitch relaxation process, thus in the relaxation
timescale, the spin down rate will gradually go back to the value predicted by fitting to pre-glitch
data. However, this is not the case for the young Crab pulsar.
Despite its very low glitch activity~(Fuentes et al. 2017), the Crab pulsar can not go back to
predicted spin down rate even till three years later after glitches~(Lyne et al. 2015), most evident
in large Crab glitch recovery processes, this phenomenon is called the persistent shift. The persistent
shift is accumulative if the time interval is less than three years and their effects
can not be resolved. Besides, several large glitches in the Crab pulsar have experienced
slow increase in spin frequency with timescales of days following the rapid rise, which mean day-long
timescale positive or at least effective positive torques, this phenomenon is called delayed spin-up.
Delayed spin-up was first discovered in the comparatively large glitch in 1989~(Lyne et al. 1992),
and in two further glitches in 1996~(Wong et al. 2001) and 2017~(Shaw et al. 2018), Table(2)
gives parameters of these three glitches. Remarkably, large Crab glitches are accompanied by both delayed
spin-ups and persistent shifts, besides, larger glitch size corresponds to longer delayed spin-up
timescale and larger persistent shift, from this point of view, it seems that delayed spin-up and
persistent shift are tightly correlated. Other young neutron star also experience persistent shift,
for instance,  PSR B2334+61 (characteristic age $\tau_{c}\sim 41~\rm{kyr}$) experienced a very large
glitch (glitch size $\Delta\nu/\nu\sim 20.5\times 10^{-6}$, much larger than Crab glitches) between
MJDs 53608 and 53621, this glitch resulted in a large long-term persistent shift amounts to $\sim1.1\%$
of the spin down rate at the time of the glitch~(Yuan et al. 2010), but no delayed spin-up is reported,
probably indicating a different physical origin.

The anomalous post-glitch behaviors of the Crab pulsar pose challenges to the standard vortex creep theory.
Alpar et al. had explained this by combining the vortex creep and starquake~(Alpar et al. 1994; Alpar
et al. 1996). They proposed that starquake would result in vortex depletion region in the crust, when
glitch is triggered and large amount of superfluid vortices move outward, part of the flowing vortices
would transport inward and be trapped by vortex depletion region, resulting in the delayed spin-up.
Besides, they interpreted the persistent shift as a decrease in effective moment of inertia through
creation of new vortex depletion regions. The differences between the Crab and the Vela pulsar are
understood from the view of evolutionary, as no new depletion region can be formed in the Vela pulsar
because it is much older than the Crab pulsar. This theory phenomenologically explain the
observations, but it depends strongly on the assumed notion of vortex depletion region that can not
be verified. Besides, within this model, the delayed spin-up and persistent shift result from
different physical origins, thus it is hard to build up any direct correlations between observables
in these two phenomena. Haskell et al. emphasized the effect of vortex accumulation and proposed
that vortex accumulation at certain part of the neutron star may account for the delayed
spin-up that is seen as a extension of the fast spin-up~(Haskell et al. 2018), but they
provided no explanation for the physical origin of persistent shift phenomenon.

This letter aims at the data analysis to infer the possible correlations between observables
of delayed spin-up and persistent shift phenomena from the view of statistics. We present the
detailed statistics and analysis in Section 2, and the summary and discussion in Section 3.

\section{statistics and analysis}
\begin{table*} \centering \caption{Observable of all Crab glitches}
\renewcommand{\arraystretch}{1.7}
\begin{tabular}{llllllllllllll}
\hline\hline
&$\rm{Date}$                &$\rm{MJD}~(\rm{d})$                              &$\Delta\nu/\nu~(10^{-9})$
&$\Delta\nu~(\mu\rm{Hz})$   &$\Delta\dot{\nu}/\dot{\nu}~(10^{-3})$   &$\Delta\dot{\nu}_{p}~(10^{-15}~\rm{s^{-2}})$  \\
\hline
&1969 September      &$40491.84(3)$       &$7.2(4)$       &$0.22(1)$        &$0.44(4)$       &                      \\
\hline
&1971 July           &$41161.98(4)$       &$1.9(1)$       &$0.057(4)$       &$0.17(1)$       &                      \\
\hline
&1971 October        &$41250.32(1)$       &$2.1(1)$       &$0.062(3)$       &$0.11(1)$       &                      \\
\hline
&1975 February       &$42447.26(4)$       &$35.7(3)$      &$1.08(1)$        &$1.6(1)$        &$-112(2)$             \\
\hline
&1986 August         &$46663.69(3)$       &$6.0$          &$0.18(2)$        &$0.5(1)$        &                      \\
\hline
&1989 August         &$47767.504(3)$      &$81.0(4)$      &$2.43(1)$        &$3.4(1)$        &$-150(5)$             \\
\hline
&1992 November       &$48945.6(1)$        &$4.2(2)$       &$0.13(1)$        &$0.32(3)$       &                      \\
\hline
&1995 October        &$50020.04(2)$       &$2.1(1)$       &$0.063(2)$       &$0.20(1)$       &                      \\
\hline
&1996 June           &$50260.031(4)$      &$31.9(1)$      &$0.953(4)$       &$1.73(3)$       &                      \\
&1997 January        &$50458.94(3)$       &$6.1(4)$       &$0.18(1)$        &$1.1(1)$        &$-116(5)$             \\
\hline
&1997 December       &$50812.59(1)$       &$6.2(2)$       &$0.19(1)$        &$0.62(4)$       &                      \\
\hline
&1999 October        &$51452.02(1)$       &$6.8(2)$       &$0.20(1)$        &$0.7(1)$        &$-25(3)$              \\
\hline
&2000 July           &$51740.656(2)$      &$25.1(3)$      &$0.75(1)$        &$2.9(1)$        &                      \\
&2000 September      &$51804.75(2)$       &$3.5(1)$       &$0.105(3)$       &$0.53(3)$       &$-53(3)$              \\
\hline
&2001 June           &$52084.072(1)$      &$22.6(1)$      &$0.675(3)$       &$2.07(3)$       &                      \\
&2001 October        &$52146.7580(3)$     &$8.87(5)$      &$0.265(1)$       &$0.57(1)$       &$-70(10)$             \\
\hline
&2002 August         &$52498.257(2)$      &$3.4(1)$       &$0.101(2)$       &$0.70(2)$       &                      \\
&2002 September      &$52587.20(1)$       &$1.7(1)$       &$0.050(3)$       &$0.5(1)$        &$-8(2)$               \\
\hline
&2004 March          &$53067.0780(2)$     &$214(1)$       &$6.37(2)$        &$6.2(2)$        &                      \\
&2004 September      &$53254.109(2)$      &$4.9(1)$       &$0.145(3)$       &$0.2(1)$        &                      \\
&2004 November       &$53331.17(1)$       &$2.8(2)$       &$0.08(1)$        &$0.7(1)$        &$-250(20)$            \\
\hline
&2006 August         &$53970.1900(3)$     &$21.8(2)$      &$0.65(1)$        &$3.1(1)$        &$-30(5)$              \\
\hline
&2008 April          &$54580.38(1)$       &$4.7(1)$       &$0.140(4)$       &$0.2(1)$        &                      \\
\hline
&2011 November       &$55875.5(1)$        &$49.2(3)$      &$1.46(1)$        &X               &$-132(5)$             \\
\hline
&2017 March          &$57839.92(6)$       &$2.14(11)$     &$0.064(3)$       &$0.27(3)$       &                      \\
&2017 November       &$58064.555(3)$      &$516.37(10)$   &$15.304(9)$    &$6.969(21)$     &                      \\
&2018 April          &$58237.357(5)$      &$4.08(22)$     &$0.122(6)$       &$0.46(11)$      &                      \\
\hline\hline
\end{tabular}\label{table1}
\end{table*}

All measured values of Crab pulsars glitches are presented in Table(1), data are taken from
references Espinoza et al. 2011 and 2014, Lyne et al. 1992, Wong et al. 2001, Shaw et al. 2018
and from website http://www.jb.man.ac.uk/pulsar/glitches/gTable.html. The first and second
columns correspond to time of the glitches, the third column is the fractional increase
in spin frequency ($\Delta\nu/\nu$), the fourth column is the step increase in spin
frequency ($\Delta\nu$), namely, glitch size,
the fifth column is the fractional increase in spin down rate ($\Delta\dot{\nu}/\dot{\nu}$),
and the last column is the persistent shift value ($\Delta\dot{\nu}_{p}$), X means unknown.
Isolated glitches are separated from each other by lines in Table(1), but neighboring glitches
whose effects are unresolved are not separated. Observables of three large glitches where
both delayed spin-up and persistent shift occurred are listed in Table(2), $\Delta\nu_{d}$
is the frequency increase in the delayed spin-up process and $\tau_{d}$ is the timescale of
delayed spin-up. Numbers in brackets represent error bars of the last significant digit.

\begin{figure}
\centering\resizebox{\hsize}{!}{\includegraphics{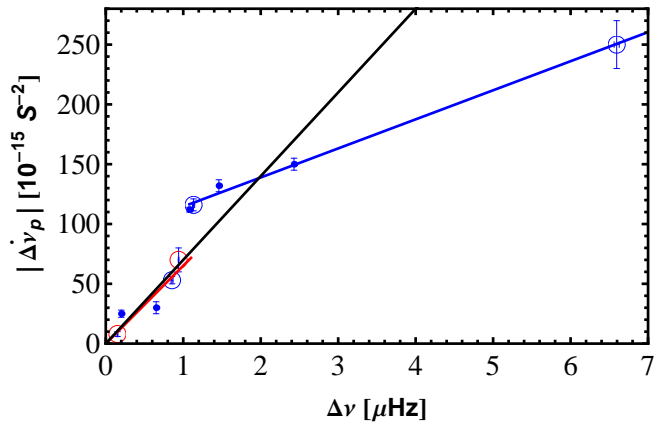}}
\caption{Persistent shift as a function of glitch size. Filled circles represent persistent
shifts in isolated Crab glitches, blue and red unfilled circles represent cumulative
persistent shifts after several neighboring glitches, but blue unfilled circles include a primary
glitch which contributes most of the persistent shift, for example, Crab glitch MJD 53067.0780,
red unfilled circles include several glitches with similar glitch sizes, for example, Crab glitch
MJD 52498.257 and MJD 52587.20. Red and blue thick lines are our linear fitting to small and
relatively large glitches separately. For comparison, Lyne's linear fitting is shown as the
black thick line. For all unfilled circles, their glitch sizes are simply the sum of glitch size
of several neighboring glitches.}
\end{figure}

\begin{figure}
\centering\resizebox{\hsize}{!}{\includegraphics{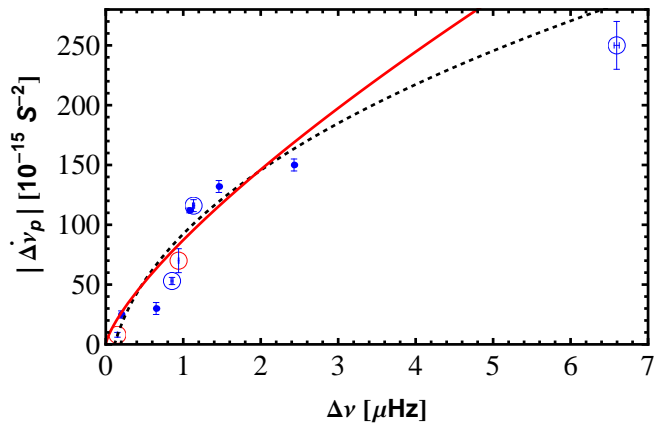}}
\caption{Same with Figure 1 but persistent shift is fitted as power law functions of glitch size.
Red thick and black dotted lines represent the fitting functions $|\Delta\dot{\nu}_{p}|=(a_{1}
\Delta\nu^{b_{1}}+c_{1})\times10^{-15}{~\rm{
s^{-2}}}$ with $c_{1}=0$ set by hand and $|\Delta\dot{\nu}_{p}|=(a_{2}\Delta\nu^{b_{2}}+c_{2})\times10^{-15}{~\rm{
s^{-2}}}$ respectively.}
\end{figure}

\begin{figure}
\centering\resizebox{\hsize}{!}{\includegraphics{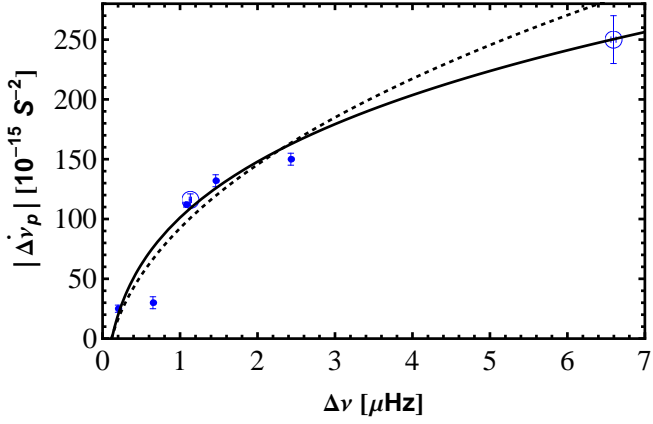}}
\caption{Same with Figure 1 but only persistent shifts in relatively isolated glitches are
considered, which means glitches MJD 51804.75, MJD 52146.7580 and MJD 52587.20 are excluded.
Black dotted and thick lines represent the fitting functions $|\Delta\dot{\nu}_{p}|=(a_{2}\Delta\nu^{b_{2}}+c_{2})\times10^{-15}{~\rm{
s^{-2}}}$ and $|\Delta\dot{\nu}_{p}|=(a_{3}\Delta\nu^{b_{3}}+c_{3})\times10^{-15}{~\rm{
s^{-2}}}$ respectively.}
\end{figure}

\begin{figure}
\centering\resizebox{\hsize}{!}{\includegraphics{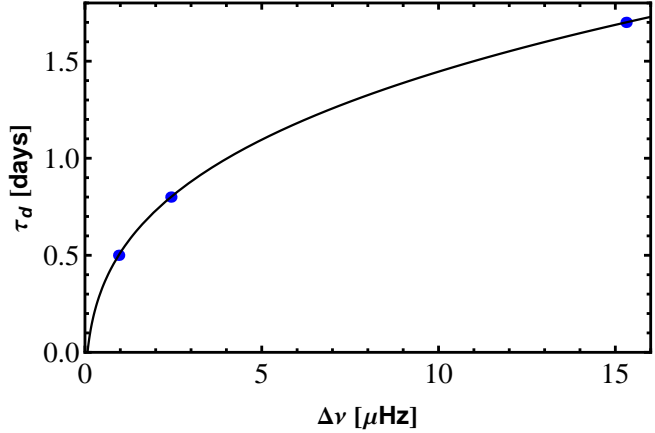}}
\caption{Delayed spin-up timescale as power law functions of glitch size. Black line represents the fitting function $\tau_{d}/({\rm{days}})
=a_{4}\Delta\nu^{b_{4}}+c_{4}$.}
\end{figure}

\begin{table*} \centering \caption{Observable of three large Crab glitches}
\renewcommand{\arraystretch}{2.0}
\begin{tabular}{llllllllllllll}
\hline\hline
&$\rm{MJD}~(\rm{d})$    &$\Delta\nu~(\mu\rm{Hz})$  &$\Delta\nu_{d}~(\mu\rm{Hz})$
&$\tau_{d}~(\rm{days})$   &$\Delta\dot{\nu}_{p}~(10^{-15}~\rm{s^{-2}})$         \\
&$47767.504(3)$   &$2.43(1)$       &$0.7$       &$0.8$    &$-150(5)$            \\
&$50260.031(4)$   &$0.953(4)$      &$0.31$      &$0.5$    &$-116(5)$            \\
&$58064.555(3)$   &$15.304(9)$     &$1.1$       &$1.7$    &X                    \\
\hline\hline
\end{tabular}\label{table2}
\end{table*}

Firstly, we analyze the relationship between $\Delta\dot{\nu}_{p}$ and $\Delta\nu$, as shown
in Figure 1, 2 and 3. Figure 1 suggests two groups of $|\Delta\dot{\nu}_{p}|$ at cutoff $\Delta\nu
\sim 1~\mu\rm{Hz}$, in the following, glitches with $\Delta\nu>1~\mu\rm{Hz}$ are called large
glitches, on the contrary, glitches with $\Delta\nu<1~\mu\rm{Hz}$ as small glitches.
A linear fitting to five large glitches gives
\begin{equation}
|\Delta\dot{\nu}_{p}|=(24\Delta\nu+90)\times10^{-15}~\rm{s^{-2}},
\end{equation}
where $\Delta\nu$ is in $\mu\rm{Hz}$ throughout this writing. While linear fitting to small
glitches gives
\begin{equation}
|\Delta\dot{\nu}_{p}|=65\times 10^{-15}\Delta\nu~\rm{s^{-2}},
\end{equation}
Lyne et al. have also considered the inter-dependence of $\Delta\dot{\nu}_{p}$ and $\Delta\nu$~
(Lyne et al. 2015), their linear fitting to all glitches gives,
\begin{equation}
|\Delta\dot{\nu}_{p}|=70\times 10^{-15}\Delta\nu~\rm{s^{-2}}.
\end{equation}
This result is very close to our fitting to small Crab glitches. A comparison between
Figure 1 in this paper and Figure 5 in Lyne et al. 2015 shows clearly that, the fitting in
logarithmic coordinate space seriously underestimated the contribution of large glitches.

Different linear relations between $|\Delta\dot{\nu}_{p}|$ and $\Delta\nu$ for large
and small Crab glitches probably indicates their differences in physical origins, for
example, large glitches may have the potential to influence internal structure of
neutron stars and result in relatively large persistent shifts, while effects of small
glitches is limited and it seems impossible to change the structure, in this case,
persistent shifts in small glitches may originate from some other unknown mechanism.
However, conclusion that persistent shifts in small and large glitches arise
from different physical processes seems to be unconvincing because of the absence of
more data. Besides, distribution of points in the $|\Delta\dot{\nu}_{p}|$ versus $\Delta\nu$
plot also influence our judgement.

\begin{figure}
\centering\resizebox{\hsize}{!}{\includegraphics{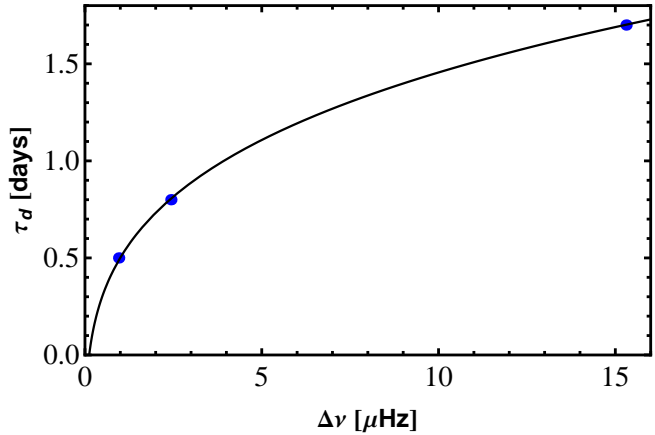}}
\caption{Application of the best fitting function of persistent shift versus glitch size to
the delayed spin-up timescale with constant $K_{1}=201.98\times10^{-15}~{\rm{s^{-2}~days^{-1}}}$.
}
\end{figure}

If the large and small glitches in the Crab pulsar do have the same physical origin,
it is reasonable to fit the whole range of data with one single function. Despite the
linear fitting we discussed above, it is naturally to consider the power law fitting
since persistent shifts of larger Crab pulsar glitches tend to be much larger.
A fitting function in
the form of $|\Delta\dot{\nu}_{p}|=a_{1}\Delta\nu^{b_{1}}\times10^{-15}{~\rm{
s^{-2}}}$ gives $a_{1}=87.24\pm1.288$, $b_{1}=0.7438\pm0.0186$, $c_{1}=0$ and $\chi^{2}/{\rm{dof}}=
304.9/8$, while a fitting function in the form of $|\Delta\dot{\nu}_{p}|=(a_{2}\Delta\nu
^{b_{2}}+c_{2})\times10^{-15}{~\rm{s^{-2}}}$ gives $a_{2}=151.5\pm13.1$, $b_{2}=0.4337
\pm0.0444$, $c_{2}=-59.08\pm12.33$ and $\chi^{2}/{\rm{dof}}=222.8/7$. The coefficient $c_{1}=0$ is obviously set by hand and the latter function gives a better fitting mathematically. Although the power law function seems to reconcile both large and small
glitches, accumulative effect of persistent shift may contaminate the fitting. Therefore,
we further
fit persistent shifts of relatively isolated glitches after exclusion of glitch MJD 51804.75,
MJD 52146.7580 and MJD 52587.20, a fitting function in the form of $|\Delta\dot{
\nu}_{p}|=(a_{3}\Delta\nu^{b_{3}}+c_{3})\times10^{-15}{~\rm{s^{-2}}}$ gives $a_{3}=243.1
\pm60.34$, $b_{3}=0.2536\pm0.06654$, $c_{3}=-141.9\pm59.51$ and $\chi^{2}/{\rm{dof}}=
102.6/4$, this have improved the fitting mathematically. Comparison between $|\Delta\dot{\nu}_{p}|=(a_{2}\Delta\nu^{b_
{2}}+c_{2})\times10^{-15}{~\rm{s^{-2}}}$ (black dotted line in Figure 3) and $|\Delta\dot{\nu}_
{p}|=(a_{3}\Delta\nu^{b_{3}}+c_{3})\times10^{-15}{~\rm{s^{-2}}}$ (black thick line in Figure 3)
shows that, $|\Delta\dot{\nu}_{p}|=(a_{3}\Delta\nu^{b_{3}}+c_{3})\times10^{-15}{~\rm{s^{-2}}}$
fits the data better while $|\Delta\dot{\nu}_{p}|=(a_{2}\Delta\nu^{b_{2}}+c_{2})\times10^
{-15}{~\rm{s^{-2}}}$ underestimates contribution of large glitch MJD 53067.0780.
Following this procedure, the function  $|\Delta\dot{\nu}_{p}|=(a_{3}\Delta\nu^{b_{3}}+c_{3})\times10^{-15}{~\rm{s^{-2}}}$
can be seen as the best fitting result.

\begin{figure}
\centering\resizebox{\hsize}{!}{\includegraphics{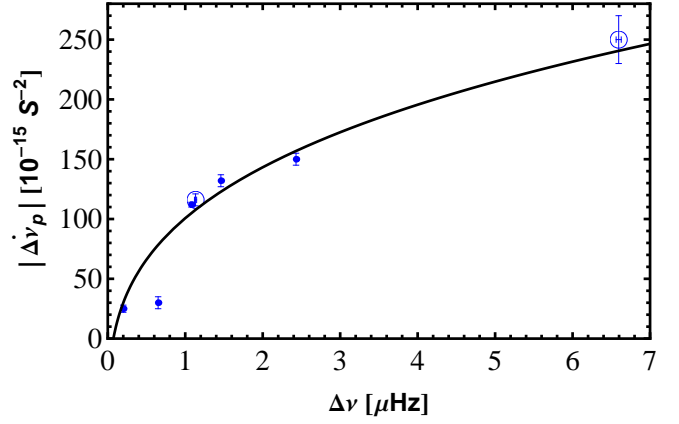}}
\caption{Application of the best fitting function of delayed spin-up timescale versus glitch
size to the persistent shift with constant $K_{2}=196.08\times10^{-15}~{\rm{s^{-2}~days^{-1}}}$.}
\end{figure}

We emphasize that, uncertainties of persistent shift in small glitches are relatively larger than that in
large glitches, which means that persistent shift in large isolated glitches are more reliable, therefore, the best fitting should be close to points with large persistent shift.
We thus removed the data points with serious accumulative effect so as not to affect the fitting. Besides, the best fitting function will inevitably bring in the cut-off glitch-size below which no persistent shift value is measured, physical meaning of this is unclear at present, probably related
to some non-ideal effects, for example, nonspherically symmetric neutron star structure.

Secondly, the three large glitches in Table (2) brings another observable, the
delayed spin-up timescale $\tau_{d}$. As all three delayed spin-ups are observed in
relatively large Crab glitches and larger glitch corresponds to longer delayed spin-up
timescale, it is natural and meaningful to consider the link between $\tau_{d}$ and $\Delta\nu$.
We perform the pure mathematical power law fitting as shown in Figure 4. A fitting function in the form of $\tau_{d}/({\rm{days}})
=a_{4}\Delta\nu^{b_{4}}+c_{4}$ (black line in Figure 4) gives $a_{4}=0.9467\pm0.022$, $b_{4}=0.2978\pm0.0046$,
$c_{4}=-0.4332\pm0.022$ with $\chi^{2}/{\rm{dof}}=4.81\times10^{-7}/0$. Though the fitting
curve goes across the data points and seems to fit the data well,
it is not reliable in principle as the data is much too less and any other
functions may fit these three points well. However, we noticed that the index $b_{4}=0.2978\pm0.0046$
is well within the uncertainty of $b_{3}=0.2536\pm0.06654$, it is probably that they are highly
identical. Using $|\Delta\dot{\nu}_{p}|=(a_{3}
\Delta\nu^{b_{3}}+c_{3})\times10^{-15}{~\rm{s^{-2}}}$ and $\tau_{d}
=(a_{4}\Delta\nu^{b_{4}}+c_{4})~{\rm{days}}$, we then try to fit
$\tau_{d}$ versus $\Delta\nu$ by $\tau_{d}=|\Delta\dot{\nu}_{p}|/K_{1}$
through minimizing the $\chi^{2}$ value, our calculations give the values $K_{1}=201.98\times
10^{-15}~{\rm{s^{-2}~days^{-1}}}$ and
$\chi^{2}=0.00021$, as shown in Figure 5. On the other hand, we try to fit $|\Delta\dot{\nu}_
{p}|$ versus $\Delta\nu$ by $|\Delta\dot{\nu}_{p}|=K_{2}\tau_{d}$ in the same way, our calculations
give $K_{2}=196.08\times10^{-15}~{\rm{s^{-2}~days^{-1}}}$ and $\chi^{2}=
2698.87$, as shown in Figure 6. This cross check shows the possibility that the persistent shifts
and delayed spin-up timescales versus glitch size follow the same power law distribution.
Furthermore, ratio of absolute persistent shift to delayed spin-up timescale for glitch MJD
47767.504 is $|\Delta\dot{\nu}_{p}|/\tau_{d}\approx 187.5\times10^{-15}~{\rm{s^{-2}~days^{-1}}}$,
and for glitch MJD 50260.031 $|\Delta\dot{\nu}_{p}|/\tau_{d}
\approx 232\times10^{-15}~{\rm{s^{-2}~days^{-1}}}$, it is obvious that $187.5\times10^{-15}~
{\rm{s^{-2}~days^{-1}}}<K_{1}\approx K_{2}<232\times10^{-15}~{\rm{s^{-2}~days^{-1}}}$, which
suggests a possible linear correlation between the persistent shift value and the delayed
spin-up timescale, the possible linear relationship can be further tested by future
measurement of the persistent shift of glitch MJD 58064.555.

Finally, we analyze the correlation between $\Delta\dot{\nu}_{p}$ and $\Delta\dot{\nu}/\dot{\nu}$,
as shown in Figure 7. The distribution is sparse and even worse if the only point in the top right
corner of Figure 7 is not considered. The sparse distribution indicates weak correlation between
$\Delta\dot{\nu}_{p}$ and $\Delta\dot{\nu}/\dot{\nu}$, which means small possibility that persistent
shift results from the decoupled moment of inertia that do not re-couple again. This is consistent
with the absence of such persistent shift in the Vela pulsar and suggests some unknown physical
difference between the Crab and Vela pulsars.

\begin{figure}
\centering\resizebox{\hsize}{!}{\includegraphics{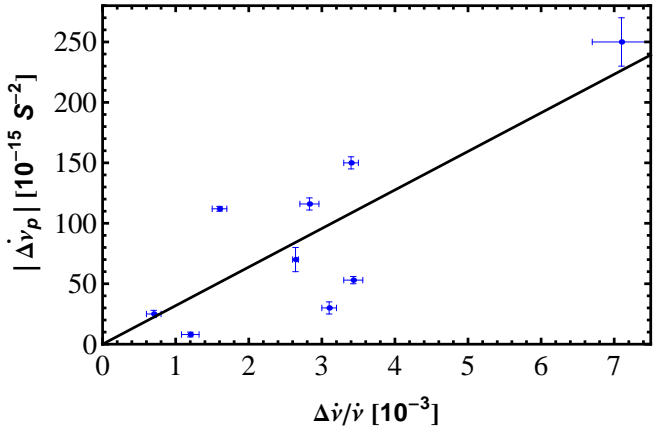}}
\caption{Persistent shift as a function of fractional increase in spin down rate. If cumulative
effect should be considered, we take the sum of fractional increase in spin down of several
neighboring glitches as the abscissa.}
\end{figure}

\section{summary and discussion}
We have performed a statistic study on all measured values during the
post-glitch recovery process in the Crab pulsar. Our pure mathematical
fitting results show that, persistent shifts for the relatively large
and small glitches may either have different linear dependence on glitch
size or follow one single power law function.

Interestingly, the fact that the power law fitting also applies to delayed
spin-up timescales demonstrates the merit of a single power law fitting to
persistent shift values. To overcome the drawback of too little delayed
spin-up timescale data, we perform the cross check by applying the
best fitting functions of persistent shift and delayed spin-up timescale
to the other one respectively, at certain coefficients $K_{1}$ and $K_{2}$
which minimize the $\chi^{2}$ values. The result that coefficients $K_{1}$
and $K_{2}$ are pretty close indicates a tight linear relationship between
the persistent shift values and the delayed spin-up timescales.
This strongly support the conclusion that they may have the same physical origin.
As an explanation,
we can expect a physical
mechanism which can result in the extra angular momentum transfer $I\Delta\nu_{d}=
\Delta I (\nu_{s}-\nu)$ and simultaneously the change in neutron star structure
denoted by the fractional change of effective moment of inertia $\Delta I/I$ through the material transfer from the
superfluid to normal component, where $\nu_{s}$ and $\nu$ are spin frequencies of
the superfluid and normal components respectively, $\nu_{s}-\nu$ is the spin lag.
The net spin-down rate of the star(the persistent shift) then arrives at $\Delta
\dot{\nu}_{p}=\dot{\nu}\Delta I/I$ by angular momentum conservation,
$\dot{\nu}$ is the spin down rate of the Crab pulsar, at present, $\nu=29.63~\rm{Hz}$
and $\dot{\nu}=-3.68\times 10^{-10}~\rm{Hz~s^{-1}}$. The Combination with the
extra angular momentum transfer immediately gives a linear relationship $\Delta
\dot{\nu}_{p}=\frac{\dot{\nu}\Delta\dot{\nu}_{d}}{\nu_{s}-\nu}\tau_{d}$, the
coefficient $\dot{\nu}\Delta\dot{\nu}_{d}/(\nu_{s}-\nu)$ has the same unit with
$K_{1}$ and $K_{2}$.  If we take the spin lag as $1.2\times10^{-3}~\rm{s^{-1}}$~
(Haskell \& Melatos 2015),
absolute value of this coefficient is about $200\times 10^{-15}~{\rm{s^{-2}~days^{-1}}}$,
supporting the above fitting results.
Monitoring of post-glitch evolution has been
applied to constrain quantities such as the fractional moment of inertia involved
in the re-coupling process and the mutual friction parameters which govern the
re-coupling between the superfluid and normal components, however, requirement
of self-consistency between the delayed spin-up and persistent shift phenomena
may set more stringent constraints on these. A new window may be opened to probe
the interior of neutron star, allowing stringently constraints on the vortex
motion, even on the nuclear equation of state in high densities.

Future measurement of persistent shift of glitch MJD 58064.555 can serve as a test to
the linear relationship between the persistent shift values and the delayed spin-up
timescales. Lyne's linear fitting predicts $\Delta\dot{\nu}_{p}\sim -1000\times 10^
{-15}~{\rm{s^{-2}}}$, while our linear fitting gives $\Delta\dot{\nu}_{p}\sim -457
\times 10^{-15}~{\rm{s^{-2}}}$, our best power law fitting gives $\Delta\dot{\nu}
_{p}\sim -(344\pm120)\times 10^{-15}~{\rm{s^{-2}}}$. Zhang et al. observed this
glitch in the 0.5-10~{\rm{keV}} X-ray band with the \emph{X-Ray Pulsar Navigation-I}
(XPNAV-1) satellite, using the first 100~\rm{days}-long post-glitch data, their
fittings gave a persistent shift $\Delta\dot{\nu}_{p}\sim-(1040\pm150)\times 10^{-15}
~{\rm{s^{-2}}}$~(Zhang et al. 2018). However, the recovery process
was not completed at that time, if the fitting function $\delta\dot{\nu}=|\Delta\dot{\nu}_{p}|
\times(0.46\times {\rm{exp}}(-t/320)-1.0)$ ($t$ is the time since the glitch epoch in units
of days) is universal for all Crab glitches, the inferred final persistent shift should be
$\Delta\dot{\nu}_{p}\sim-(1567\pm226)\times 10^{-15}~{\rm{s^{-2}}}$.

It should be noticed that, fittings in section 2 are relatively rough at present
because of (i)the lack of more data points, (ii)the effect of contamination of
neighboring glitches, (iii)the non-uniform distribution of data points in glitch
size. Thus, more persistent shift and delayed spin-up events are urgently needed
for statistics and theoretical work.

\acknowledgements The authors would like to thank Jin Wu from Key Laboratory of
Quark and Lepton Physics (MOE) and Institute of Particle Physics, Central China
Normal University for his help in data analysis. This work is supported by
National Natural Science Foundation of China (Grant No. 11773011). Correspondence
should be addressed to W. H. Wang (email: wangweihua@mails.ccnu.edu.cn) and
X. P. Zheng (email: zhxp@phy.ccnu.edu.cn).


\begin{thebibliography}{99}
\bibitem[Taylor et al.(1993)]{}Taylor, J. H., Manchester, R. N., Lyne, A. G. 1993, VizieR Online Data Catalog, 7156
\bibitem[Radhakrishnan \& Manchester (1969)]{1969Nature...222...228}Radhakrishnan, V., \& Manchester, R. N. 1969, Nature(London), 222, 228
\bibitem[Reichely \& Downs (1969)]{1969Nature...222...229}Reichely, P. E., \& Downs, G. S. 1969, Nature(London), 222, 229
\bibitem[Ruderman (1969)]{1969Nature...223...597}Ruderman, M. 1969, Nature(London), 223, 597
\bibitem[Pines et al. (1972)]{1972Nature Physical Science...237...83}Pines, D., Shaham, J., Ruderman, M. 1972, Nature Phys. Sci., 237, 83
\bibitem[Michel (1970)]{1970ApJL...159...L25}Michel, F. C. 1970, ApJL, 159, L25
\bibitem[Scargle \& Pacini (1971)]{1971Nature Physical Science...232...144}Scargle, J. D., \& Pacini, F. 1971, Nature Physical Science, 232, 144
\bibitem[Pines et al. (1974)]{}Pines, D., Shaham, J., Ruderman, M., 1974, IAU proceedings, 53, 189
\bibitem[Baym et al. (1969)]{1969Nature...224...673}Baym, G., Pethick, C., Pines, D., 1969, Nature(London), 224, 673
\bibitem[Anderson \& Itoh (1975)]{1975Nature...256...25}Anderson, P. W., \& Itoh, N., 1975, Nature(London), 256, 25
\bibitem[Alpar et al. (1984)]{1984ApJ...276...325}Alpar, M. A., Anderson, P. W., Pines, D., \& Shaham, J., 1984, ApJ, 276, 325

\bibitem[Fuentes et al. (2017)]{2017AA...608...131}Fuentes, J. R., Espinoza, C. M., Reisenegger, A., et al. 2017, A\&A, 608, 131
\bibitem[Lyne et al. (2015)]{2015MNRAS...446...857}Lyne, A. G., Jordan, C. A., Graham-Smith, F., et al. 2015, MNRAS, 446, 857
\bibitem[Lyne et al. (1992)]{1992Nature...359...706}Lyne, A. G., Graham-Smith, F., \& Pritchard, R. S., 1992, Nature, 359, 706
\bibitem[Wong et al. (2001)]{2001ApJ...548...447}Wong, T., Backer, D. C., Lyne, A. G., 2001, ApJ, 548, 447
\bibitem[Shaw et al. (2018)]{2018MNRAS...478...3832}Shaw, B., Lyne, A. G., Stappers, B. W., et al. 2018, MNRAS, 478, 3832
\bibitem[Yuan et al. (2010)]{2010ApJL...719...L111}Yuan, J. P., Manchester, R. N., Wang, N., et al. 2010, ApJL, 719, L111


\bibitem[Alpar et al. (1994)]{1994ApJ...427...L29}Alpar, M. A., Chau, H. F., Cheng, K. S., et al. 1994, ApJ, 427, L29


\bibitem[Espinoza et al. (2011)]{2011MNRAS...414...1679}Espinoza, C. M., Lyne, A. G., Stappers, B. W., et al. 2011, MNRAS, 414, 1679
\bibitem[Espinoza et al. (2014)]{2014MNRAS...440...2755}Espinoza, C. M., Antonopoulou, D., Stappers, B. W., et al. 2014, MNRAS, 440, 2755


\bibitem[Haskell \& Melatos (2015)]{}Haskell, B., \& Melatos, A., 2015, International Journal of Modern Physics D, 24, 1530008
\bibitem[Zhang et al. (2018)]{2018ApJ...866...82}Zhang, X. Y., Shuai, P., Huang, L. W., et al. 2018, ApJ, 866, 82



\end{thebibliography}
\end{document}